\documentclass[sigconf]{acmart}

\copyrightyear{2020}
\acmYear{2020}
\setcopyright{acmcopyright}\acmConference[MM '20]{Proceedings of the 28th ACM International Conference on Multimedia}{October 12--16, 2020}{Seattle, WA, USA}
\acmBooktitle{Proceedings of the 28th ACM International Conference on Multimedia (MM '20), October 12--16, 2020, Seattle, WA, USA}
\acmPrice{15.00}
\acmDOI{10.1145/3394171.3413848}
\acmISBN{978-1-4503-7988-5/20/10}

\begin{document}
\fancyhead{}
\title{Temporally Guided Music-to-Body-Movement Generation}








\author{Hsuan-Kai Kao and Li Su}
\affiliation{%
  \institution{Institute of Information Science, Academia Sinica, Taiwan}
  \streetaddress{128 Academia Road, Section 2, Nankang, Taipei 115, Taiwan}}
\email{{hsuankai, lisu}@iis.sinica.edu.tw}



\begin{abstract}
  This paper presents a neural network model to generate virtual violinist’s 3-D skeleton movements from music audio. Improved from the conventional recurrent neural network models for generating 2-D skeleton data in previous works,  the proposed model incorporates an encoder-decoder architecture, as well as the self-attention mechanism to model the complicated dynamics in body movement sequences. To facilitate the optimization of self-attention model, beat tracking is applied to determine effective sizes and boundaries of the training examples. The decoder is accompanied with a refining network and a bowing attack inference mechanism to emphasize the right-hand behavior and bowing attack timing. Both objective and subjective evaluations reveal that the proposed model outperforms the state-of-the-art methods. To the best of our knowledge, this work represents the first attempt to generate 3-D violinists’ body movements considering key features in musical body movement.
\end{abstract}



\begin{teaserfigure}
  \includegraphics[width=\textwidth]{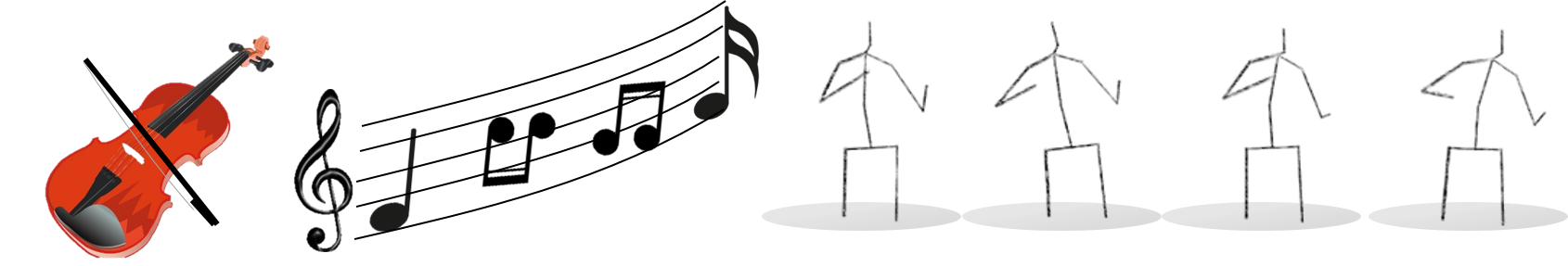}
  \caption{Our task is to create reasonable and natural playing movement with corresponding violin music.}
  \vspace{0.25cm}
  \label{fig:teaser}
\end{teaserfigure}

\maketitle

\section{Introduction}

Music performance is typically presented in both audio and visual forms. Musician's body movement acts as the pivot to connect audio and visual modalities, since musicians employ their body movement to produce the performed sound, and such movement also serves as the means to communicate their musical ideas toward the audience.
As the result, the analysis, interpretation, and modeling of musicians' body movement has been an essential research topic in the interdisciplinary fields for music training 
\cite{farber1987discovering, pierce1997four}, 
music recognition \cite{li2017audiovisual,huang2019identifying}, biomechanics, and music 
psychology \cite{davidson2012bodily,huang2017conducting,thompson2012exploring,wanderley2005musical,Burger2013PerceivedEmotions}. 
Motion capture and pose estimation techniques \cite{pavllo20193d} facilitated quantitative analysis of body motion by providing the data describing how each body joint moves with time.
Beyond such research works based on analysis, an emerging focus is to develop a generative model that can automatically generate body movements from music. Such a technique can be applied to music 
performance animation, and human-computer interaction platforms, in which the virtual character's body movement can be reconstructed from audio signal alone, without the physical presence of human musician.
Several studies endeavor to generate 
body movement from audio and music signals, including 
generating pianist's and violinist's 2-D skeletons from music audio \cite{shlizerman2018audio,li2018skeleton,liu2020body}, generating hand gestures from conversational speech \cite{ginosar2019learning}, and generating choreographic movements from music \cite{kakitsuka2016choreographic,lee2019dancing}. 

In this paper, we focus on the generation for violinists' body movement. 
Violinists' body movement is highly complicated and intertwined with the performed sound. To investigate musical movement, previous research identified three main 
types of body movement in music performance. First, the \emph{instrumental movement} leads to the generation of instrument sound; second, the \emph{expressive movement} induces visual cues of emotion and musical expressiveness; and third, the \emph{communicative movement} interacts with other musicians and the audience \cite{wanderley2005musical}. Taking a violinist's instrumental movement as an example, 
a \emph{bow stroke} is a movement executed by the right hand to make the bow moving across the string. For a bowed note termed \emph{arco}, there are two typical bowing modes: up-bow (the bow moving upward) 
and down-bow (the bow moving downward) 
The arrangement of bow strokes depends on how the 
musician segments a note sequence. In general, a group of notes marked with a \emph{slur} on the score 
should be played in one bow stroke. Yet the music scores do not usually contain detailed bowing annotations for every note through the whole music piece, but only provide suggested bowing marks for several important instances, which renders musicians a lot of freedom to apply diverse bowing strategies according to their own musical interpretations. Given the flexibility of bowing in the performance practice, still, the bowing configuration should be arranged in a sensible manner to reflect the structure in music compositions.
An unreasonable \emph{bowing attack} (i.e., the time instance when the bowing direction changes) timing can be easily sensed by 
experienced violinists. 
Likewise, 
the left-hand fingering movement is also flexible to a certain extent: an identical 
note can be played with different strings at different fingering positions, depending on the pitches of successive notes. 
In addition to the instrumental movements (bowing and fingering motion), which are directly constrained by the written note sequence in the music scores, 
the expressive body movements also reflect the context-dependent and subject-dependent musical semantics, including the configuration of 
beat, downbeat, phrasing, valence, and arousal in music 
\cite{pierce1997four, Burger2013PerceivedEmotions}. 
In sum, the musical body movements have diverse functions and are
attached to various types of music semantics, which leads to the 
high degree of freedom for movement patterns during the performance. The connection 
between the performed notes and body movements (including the right-hand bowing movements and left-hand fingering movements) is not one-to-one correspondence, but is highly mutual-dependent. 
Such characteristics not only make it difficult to model the correspondence between music and body movement, but also result in 
issues regarding the 
assessment of generative model: since there is no exact ground truth in body movement for a given music piece, it is not certain that if the audience's perceptual quality can be represented by 
simplified training objective (e.g., the distance between the predicted joint position and a joint position selected from a known performance). 





\begin{figure*}[ht]
\centering\includegraphics[width=\textwidth]{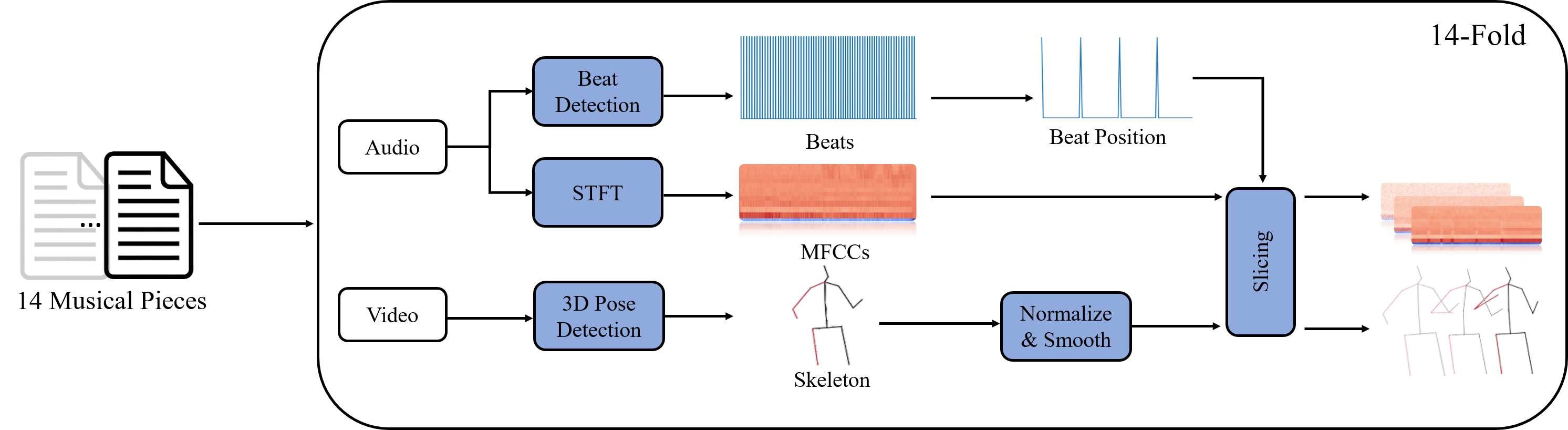}
\caption{The full process of data pre-processing.}\label{pre-processing}
\end{figure*}

In this paper, we propose a 3-D violinist's body movement generation system, which incorporates musical semantics including the beat timing 
and bowing attack inference mechanisms. Following the track 
in \cite{liu2020body}, we model the trunk 
and the right hand segments 
separately, and further develop 
this approach into an end-to-end, multi-task learning framework. To incorporate the musical semantic information  in model training, the beat tracking technique is applied to guide the processing of input data. Moreover, a state-of-the-art 3-D pose estimation technique is employed to capture the depth information of skeleton joints, which is critical in identifying bowing attacks. The pose estimation process provides reliable \emph{pseudolabels} motion data to 
facilitate the training process. To investigate the non-one-to-one motion-music correspondence, 
we propose a new dataset, which contains music with multiple performance versions by different violinists for the same set of repertoire. The generative models are evaluated on multiple performance in order to reduced the bias. To the best or our knowledge, this work represents the first attempt to generate 3-D violinists’ body movements, as well as to consider information from multiple performance versions for the development of 
body movement generation system.

The rest of this paper is organized as follows. Section 2 presents a survey of 
recent research regarding 
body movement generation techniques. The proposed method is introduced hereafter, where Section 3 describes the data processing, and Section 4 describes the model implementation. Section 5 reports the experiment and results, followed by the conclusion 
in Section 6.



\section{Related work}
\textbf{Music body movement analysis} The role of music body movement has been discussed in many studies 
\cite{davidson2012bodily}. Music performer's body movements are divided into three types: 1) the 
instrumental movement such as striking the keyboard on piano or pressing the strings on violin; 2) the expressive movement such as body swaying and head nodding; and 3) the communicative movement such as cuing movement suggesting tempo changes in music \cite{wanderley2005musical}. Studies showed that music performers could intentionally adopt different body movements to achieve the planed performance sound according to the musical context \cite{macritchie2012intentions, davidson2012bodily, macritchie2013inferring, huang2017conducting}.
For instance, different violinists may choose various bowing and fingering strategies depending on the musical interpretations they attempt to deliver. 


Previous research has shown that body movements from different music performers generate diverse instrumental sounds \cite{dahl2010gestures, macritchie2012intentions}. The correspondence between music performer' movement and the musical composition being performed has also been discussed \cite{haga2008correspondences, thompson2012exploring}. Recently, a study employs body movement data with the recurrent neural network (RNN) model to predict dynamic levels, articulation styles, and phrasing cues instructed by the orchestral conductor 
\cite{huang2019identifying}. Since detecting musical semantics from the body movement data is possible, an interesting yet challenging task is to generate body movement data from given musical sound \cite{shlizerman2018audio, li2018skeleton}. 

\textbf{Generating audio from body movement} Techniques have been developed to generate 
speech or music signals from body movement 
\cite{chung2017lip, yoon2019robots}. \cite{chung2017lip} generated human speech audio from automatic lip reading on the face videos, whereas \cite{yoon2019robots} generated co-speech movements including iconic and metaphoric gestures from speech audio. \cite{chen2017deep} applied Generative Adversarial Networks (GAN) to produce 
music performer's images 
based on different types of timbre. \cite{berg2012interactive} generated music from gathering the motion capture data. In the field of interactive multimedia, using gesture data to induce sound morphing or related generation task is also commonly used.

\textbf{Body movement generation from audio}  
Several attempts have been devoted to generate music-related movement. The commonly seen topics of body movement generation from audio include generating body movements from music, generating gestures from speech, and generating dance from music. 
\cite{shlizerman2018audio} used an RNN with long-short-term-memory (LSTM) units to encode audio features, and then employed a fully-connected (FC) layer to decode it into the body skeleton keypoints of either pianists or violinists. In \cite{kakitsuka2016choreographic}, choreographic movements are automatically generated from music according to the user's preference and the musical structural context, such as the metrical and dynamic arrangement in music. 
Another recent work on pianists' body skeleton generation \cite{li2018skeleton} also consider musical information including bar and beat positions in music. The model combining CNN and RNN was proven to be capable of learning the body movement characteristics of each pianist .



\section{Data and pre-processing}
In this section, we introduce the procedure to compile a new violin performance dataset for this study. 
And the 
data pre-processing procedure is summarized 
in Figure~\ref{pre-processing}.

\subsection{Dataset}
We propose a newly-collected 
dataset containing 140 violin solo videos with total length of around 11 hours. 
14 selected violin solo pieces were performed by 10 violin-major students from music college. 
This dataset therefore contains diverse performed version and individual musical interpretations based on the same set of repertoire, which is specifically designed for the exploration of non-one-to-one correspondence between music motion and audio.
The selected repertoire contains 12 conventional 
Western classical pieces for violin solo 
ranging from Baroque to post-Romanticism, plus two 
non-Western folk songs.

We collected 10 different versions performing identical music pieces, which allows us to derive 10 sets of bowing and fingering arrangements, as well as 
pseudolabel (i.e. the skeleton motion data extracted from pose estimation method) for each music piece. 
The multi-version design of the dataset is incorporated with our data splitting strategy to explore diverse possible motion patterns corresponding to identical music piece. The skeleton and music data are available at the project link (see Section \ref{sec:conclusion}).  

\subsection{Audio feature extraction}
We apply {\it librosa}, a Python library for music signal processing \cite{mcfee2015librosa}, to extract audio features. 
Each music track is sampled at 44.1 kHz, and the short-time Fourier transform (STFT) is performed with a sliding window (length = 4096 samples; hope size = 1/30 secs). 
Audio features are then extracted from STFT, including 
13-D Mel-Frequency Cepstral Coefficients (MFCC), logarithm mean energy (a representation for sound volume), and their first-order temporal derivative, resulting in a feature dimension of 28.



\subsection{Skeletal keypoints extraction}
The state-of-the-art pose detection method \cite{pavllo20193d} 
is adopted to extract the 3-D position of 
violinists’
15 body joints, resulting in a 45-D body joint vector for each time frame. The 15 body joints are: head, nose, thorax, spine, right shoulder, left shoulder, right elbow, left elbow, right wrist, left wrist, hip, right hip, left hip, right knee, and left knee. The joints are extracted frame-wisely at the video’s frame rate (30 fps). All the joint data are normalized, such that the mean of all joints over all time instances is zero. The normalized joint data are then smoothed over each joint using a median filter (window size = 5 frames). 

\subsection{Data pre-processing}

The extracted audio and skeletal data are 
synchronized with each other with the frame rate of 30 fps. 
To facilitate the training process, the input data are divided into segments according to the basic metrical unit in music. Beat position serves as the reference to slice data segments, considering the fact that 
the arrangement of bowing stroke is highly related to the metrical position. 
To obtain beat labels from audio recordings, we first derive 
beat positions in the MIDI file for each musical piece, and  
the dynamic time warping (DTW) algorithm is applied to align beat positions between the MIDI-synthesized audio and the recorded audio performed by human violinists. The beat positions are then used for the data segmentation. 
Each data segment starts from a beat position, and is with the length of 900, i.e., 30 seconds. According to the average tempo in the dataset, 30 seconds is slightly longer than 16 bars in music, which provides a sufficient context for our task. 
All the segmented data are normalized in feature dimension by z-score. 

For the data splitting, 
a leave-one-piece-out (i.e., 14-fold cross-validation) scheme is performed by assigning 14 pieces to the testing set by turns. 
we take the recordings of one specific violinist for training and validation, and take the recordings of the remaining nine violinists for testing. For the training and validation data, we 
choose the recordings played by the violinist whose performance technique is the best among all according to expert's opinion. Within the training and validation set, 80 \% of the sequences are for training and 20 \% of the sequences are for testing. 
This 14-fold cross-validation procedure results in 14 models. Each model is evaluated on the  
piece performed by the remaining nine violinists in 
the testing set. The results will then be discussed by comparing 
the nine different performance versions and their corresponding ground truths. This evaluation procedure can reflect the nature of violin performance, in which multiple possible motion patterns may correspond to a identical music piece in different musician's recordings.



For the cross-dataset evaluation, we also evaluate our model using the URMP dataset  \cite{li2018creating}, which has been used in previous studies for music-to-body-movement generation \cite{li2018skeleton,liu2020body}. The URMP dataset comprises 43 music performance videos 
with individual instruments recorded in separate tracks, and we choose 33 tracks containing solo violin performance as our 
test data for cross-dataset evaluation. For reproducibility, the list of the 33 chosen pieces are provided on the project link.


\begin{figure*}[t]
\centering\includegraphics[width=\textwidth]{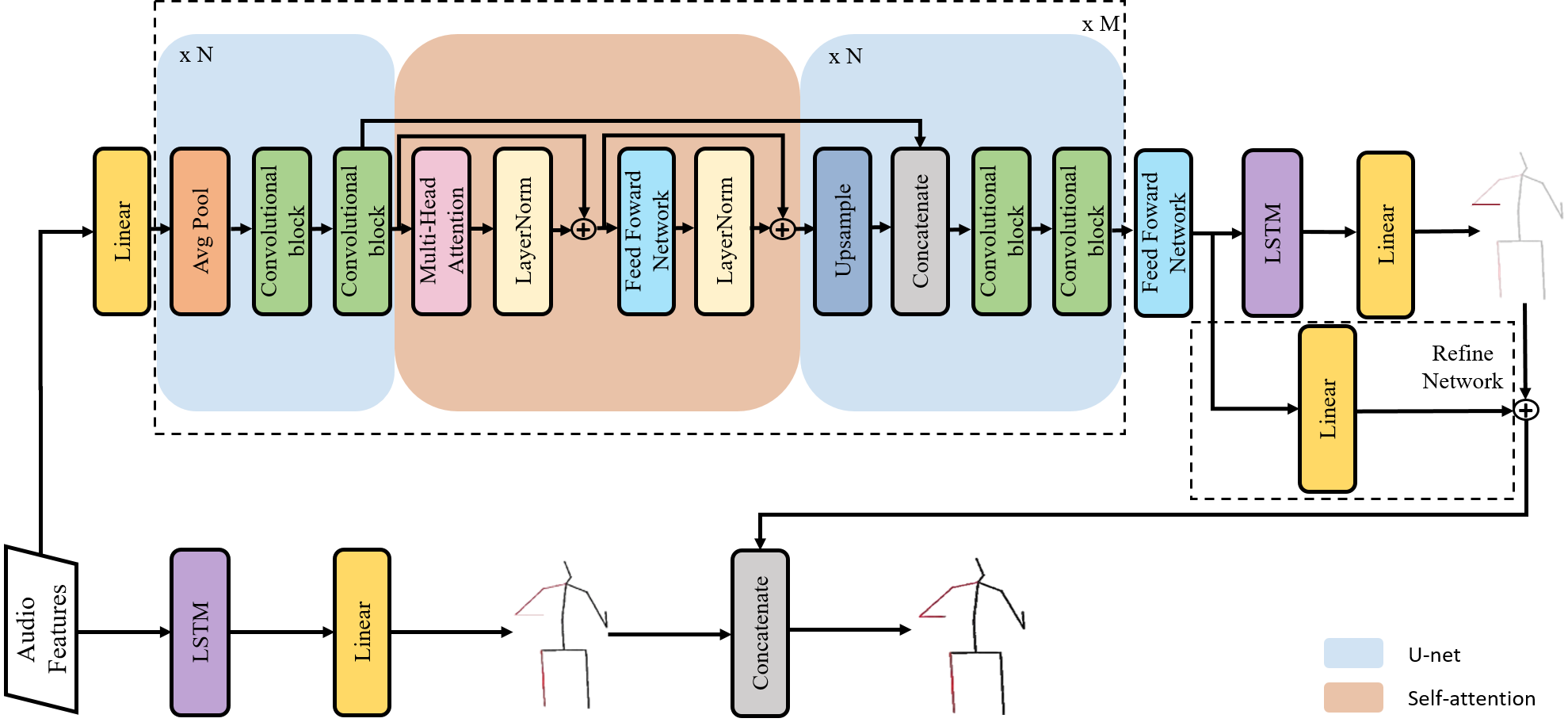}
\caption{The overview of body movement generation network.}\label{overview}
\end{figure*}

\section{Proposed model}
The 
architecture of 
proposed music-to-body-movement generation model is shown in Figure \ref{overview}. The architecture is constructed by 
two 
branches of networks: body network and right-hand network.
In order to capture the detailed variation of right-hand keypoints in the performance, the right-hand network includes one encoder and one decoder, while the body network only includes one decoder. Both networks take the audio features mentioned in Section 3.2 as the input. The feature is represented as $X:=\{x_i\}^L_{i=1}$, where $x_i\in\mathbb{R}^
28$ is the feature at the $i$th frame. In this paper, we have $L=900$. The right-hand encoder combines a U-net architecture~\cite{ronneberger2015u} with a self-attention mechanism~\cite{vaswani2017attention} at the bottleneck layer of the U-net. Based on the design of the Transformer model~\cite{vaswani2017attention}, the output of the U-net is 
fed into a position-wise feed-forward network. Its output is then fed into a recurrent model for body movement generation, which is constructed by an LSTM RNN, followed by a linear projection layer. 
The final output of the model is the superposition of the generated body skeleton $Y^{(body)}:=\{y^{(body)}_i\}^L_{i=1}$ and right-hand skeleton $Y^{(rh)}:=\{y^{(rh)}_i\}^L_{i=1}$, where $y^{(body)}_i\in\mathbb{R}^{39}$ and $y^{(rh)}_i\in\mathbb{R}^{6}$. In addition, to enhance the modeling of the right-hand joint, another linear projection layer is imposed on the right-hand wrist joint, 
and output a right-hand wrist joint calibration vector of the $y^{(rw)}_i\in\mathbb{R}^3$. This term is then added to the corresponding right-hand element of $y^{(rh)}_i$, and the right-hand decoder outputs the whole estimated right-hand skeleton. Finally, we combine the right-hand and body skeleton to output the whole estimated full skeleton $Y^{(full)}:=\{y^{(full)}_i\}^L_{i=1}$, where $y^{(full)}_i\in\mathbb{R}^{45}$. Note that our decoder mainly follows the design in \cite{shlizerman2018audio}, while our model is to indicate the significance of using a U-net-base encoder architecture with self-attention mechanism.


\subsection{U-net}
The U-net architecture~\cite{ronneberger2015u} was originally proposed to solve the image segmentation problem. Recently, it has also been widely applied to generation tasks over different data modality, due to the advantage in translating features to another modal. Examples include sketch to RGB pixel~\cite{isola2017image}, audio-to-pose generation~\cite{shlizerman2018audio}, and music transcription~\cite{wu2018automatic}. In this work, we first 
map the input features into a high-dimension through a linear layer. The output of linear layer is taken as the input of the U-net. The left part of the U-net structure starts from an average pooling layer to downsample the full sequence, and is followed by two consecutive convolutional blocks, each of which consists of one convolutional layer, 
a batch normalization layer, and ReLU activation. Such computation repeats by $N$ times until the bottleneck layer of U-net. In this paper, we set $N=4$. The main function of the encoding process of the U-net is to extract high-level features from low-level ones; 
in our scenario, it functions as a procedure to learn structural features from the frame-level audio features. The self-attention layer 
between the encoding and decoding parts of the U-net will be introduced in the next section. Next, the encoder part of the U-net starts from an upsampling layer using linear interpolation, which is concatenated with the down-sampling convolutional layer in the encoder part through the skip-connection, and then followed by two convolutional blocks. 
This calculation also repeats by $N$ times until the features are converted into another modal. Compared to the original architecture of U-net, we do not directly transform audio features to skeleton; rather, we first convert such output representation into another high-dimensional feature, which is leaved for generation task with the remaining LSTM network. Moreover, we find out that the bowing attack accuracy can be improved by stacking multiple blocks in U-net with self-attention. 
The whole block is framed by dash line, as illustrated in Figure~\ref{overview}.

\subsection{Self-attention}
Music typically has a long-term hierarchical structure. Similar patterns may appear repeatedly in a training sample. To decode the order of the body movement sequence is a critical issue. 
However, while the U-net utilizes convolutional blocks in downstream to encode audio features to symbolic representation, it merely deals with the local structure in a limited kernel size. To solve the problem of long-term sequential inference, recently, the self-attention mechanism \cite{vaswani2017attention} has been widely applied in sequence-to-sequence tasks, such as machine translation, and text-to-speech synthesis. Different from the RNN-based models, in Transformer, representation is calculated by the weighted sum of each frame of the input sequence across different states, and the more relevant states are given more weights. Accordingly, each state perceives the global information, and this would be helpful for modeling long-term information such as notes and music structure. We therefore apply the self-attention mechanism at the bottleneck layer of U-net.

\textbf{Scaled Dot-Product Attention}
Given input sequence $X \in \mathbb{R}^{L\times d}$, we first project $X$ into three matrices, namely query $Q:=XW^Q$, key $K:=XW^K$ and value $V:=XW^V$, where $W^Q, W^K, W^V \in \mathbb{R}^{d\times d}$ and $Q, K, V \in \mathbb{R}^{L\times d}$. The scaled dot-product attention computes outputs for a sequence vector $X$ as:
\begin{equation} \label{eq:1}
    \text{Attention}(Q, K, V)=\text{softmax}\left(\frac{QK^T}{\sqrt{d}}\right)V\,,
\end{equation}
where the scalar $\frac{1}{\sqrt{d}}$ is used to avoid overflowed value leading to very small gradient. 

\textbf{Multi-Head Attention}
Multi-head attention allows the model to jointly attend to the information from different representation subspaces at different positions. The scale-dot product is computed $h$ times in parallel with different \emph{head}, and the $h$th head can be expressed as follows:
\begin{equation} \label{eq:2}
    \text{Head}_h(Q_h, K_h, V_h)=\text{softmax}\left(\frac{Q_hK_h^T}{\sqrt{d_{h}}}\right)V_h\,.
\end{equation}

For each head, queries, keys, and values are projected into a subspace with dimension $d_h$, where $d_h=d/h$, $W_h^Q, W_h^K, W_h^V \in \mathbb{R}^{d\times d_h}$ and $Q_h, K_h, V_h \in \mathbb{R}^{L\times d_{h}}$. The output of each head are concatenated and linearly projected, and skip connection is applied with input $X$: 
\begin{align}
    \text{MultiHead}&=\text{Concat}(\text{Head}_1,...,\text{Head}_h)W^O\,, \\
    \text{MidLayer}&=\text{MultiHead}+X\,,
\end{align}
where $W^O \in \mathbb{R}^{(h \times d_h) \times d}$. 

\textbf{Relative Position Representations}
While there is no any positional information applied in scaled dot product, the same input at different time steps would contribute to the same attending weights. To solve the problem, we apply the relative position encoding~\cite{shaw2018self} in the scaled dot-product self-attention. Two learnable embeddings $R^K$ and $R^V$ represent the distance between two positions in sequence vector $X$, where $R^V, R^K \in \mathbb{R}^{L\times d}$, and they are shared across all attention heads. We then modify Equation~\ref{eq:1} as follows:
\begin{equation} 
\begin{aligned}
    \text{Attention}(Q, K, V)= 
    \text{softmax}\left(\frac{QK^T + Q(R^K)^T}{\sqrt{d}}\right)(V + R^V)\,.
\label{eq:5}
\end{aligned}
\end{equation}

By adding the term $Q(R^K)^T$ in numerator, the original matrix multiplication in Equation~\ref{eq:1} would be injected the relative position information. The similar way is also applied to the value term, $V + R^V$.

\textbf{Position-wise Feed Forward Network}
Another sub-layer in the self-attention block is position-wise a feed-forward network. It consists in two linear transformation layers with a ReLU activation 
between them, which is applied to each position separately and identically. The dimensionality of input and output is $d$, and the inner layer has the dimensionality of $d_{ff}$.  The outputs of this sub-layer are computed as:
\begin{equation}
    \text{FFN}(X)=\max(0,XW_1+b_1)W_2 + b_2\,,
\end{equation}
where the weights $W_1 \in \mathbb{R}^{d \times d_{ff}}$, $W_2 \in \mathbb{R}^{d_{ff} \times d}$ and biases $b_1 \in \mathbb{R}^{d_{ff}}$, $b_2 \in \mathbb{R}^{d}$. 

Additionally, we also place an extra position-wise feed forward network after the last layer of U-net. While the outputs of U-net is contextualized, the position-wise feed forward network make it more similar to the skeletal representation.


\subsection{Generation}
For the generated body sequence $\hat{Y}^{(body)}$, we directly feed audio features into the LSTM RNN network, followed by a dropout and a linear projection layer, as shown in the lower branch of Figure \ref{overview}. This branch of model directly generates the sequence of 39-D body skeleton $\hat{Y}^{(body)}$. For the right-hand sequence generation $\hat{Y}^{(rh)}$, the output of position-wise feed forward network would be fed into two components. The first one is identical to the body sequence generation network, 
and the second component is a 
network to refine the right-hand motion. While directly producing full right-hand from one branch may limits the variation of wrist joint, we take the output of position-wise feed-forward network into another branch to generate the 3-D coordinate for the right-hand wrist joint. Therefore, the right-hand output is a 6-D right-hand sequence, whose last three dimensions (represented as wrist joint) are added by the output of refine network. Finally, we concatenate the outputs of body sequence and right-hand sequence:
\begin{equation}
    \hat{Y}^{(full)} = \text{Concat}(\hat{Y}^{(body)},\hat{Y}^{(rh)})\,.
\end{equation}

The model is optimized by minimizing the $L_1$ distance between the generated skeleton $\hat{Y}^{(full)}$ and the ground truth skeleton $Y^{(full)}$: $\mathcal{L}_{\text{full}}:=\|Y^{(full)}-\hat{Y}^{(full)}\|$.

\subsection{Implementation details}
In our experiments, we use 4 convolutional blocks $(N=4)$ in the downstream and upstream subnetworks of U-net individually, and all the dimensions of convolutional layers in U-net are set to 512. In the bottleneck layer of U-net, we adopt 1 attention block. The number of head in the attention block is 4, and $d$ is set to 512. The inner dimension of feed forward network $d_{ff}$ is set to 2048. The dimension of the LSTM unit is 512, and the dropout rate for all dropout layers is 0.1. Besides, we further stack two full components $(M=2)$ composed of U-net and self-attention as our final network architecture. 

The model is optimized by Adam with $\beta_1=0.9$, $\beta_1=0.98$, $\epsilon=10^{-9}$, and adaptive learning rate is adopted over the training progress:
\begin{equation}
    lr=k\cdot d^{-0.5}\cdot min(n^{-0.5},n\cdot warmup^{-1.5}),
\end{equation}
where $n$ is the step number, $k$ is a tunable scalar. The learning rate is increased linearly for the first $warmup$ training steps and is decreased thereafter proportionally to the inverse square root of the step number. We set $warmup=500$, $k=1$ for training model with 100 epochs, and batch size is set to 32. Furthermore, we use the early-stopping scheme to choose optimal model when the validation loss stops decreasing for 5 epochs.
\begin{table*}[t]
\caption{The comparison between baselines and our proposed model in different evaluation metrics.}
\begin{tabular}{|l|cccccccc|}
\hline
Method                            & $L_1$ avg.  & $L_1$ hand avg. & PCK    & Bowx   & Bowy   & Bowz   & Bow avg. & Cosine Similarity \\ \hline\hline
{\it A2B}~\cite{shlizerman2018audio}                         & 0.0391 & 0.0925  & 0.7240  & 0.4169 & 0.4462   & 0.4062 & 0.4231  & 0.6865            \\
{\it S2G}~\cite{ginosar2019learning}                         & \textbf{0.0365} & 0.0910   & \textbf{0.7590}  & 0.3800  & 0.4330 & 0.3729 & 0.3953  & 0.6740             \\ \hline

{\it Unet1}                       & 0.0382 & 0.0870   & 0.7354 & 0.4350  & 0.4773 & 0.4130 & 0.4417  & 0.6850            \\
{\it{Unet1 + FFN + Refine Network}}                 & 0.0377 & \textbf{0.0840}   & 0.7416 & 0.4427  & 0.4870 & \textbf{0.4157} & 0.4485 & 0.6934            \\
{\it{Unet2 + FFN + Refine Network}} & 0.0379 & 0.0860   & 0.7394 & \textbf{0.4476} & \textbf{0.5165} & 0.4080 & \textbf{0.4574} & \textbf{0.6968}\\
\hline
\end{tabular}
\label{tab:quantitative results}
\end{table*}

\begin{table}[t]
\caption{The comparison between baselines and our proposed model in the URMP dataset.}
\begin{tabular}{|l|cccc|}
\hline
             & Bowx  & Bowy & Bowz  & Bow avg. \\ \hline\hline
\textit{A2B}~\cite{shlizerman2018audio} & 0.4554  & 0.4775 & 0.4448 & 0.4893  \\
\textit{S2G}~\cite{ginosar2019learning} & 0.3985 & 0.4660 & 0.4030 & 0.4225   \\ \hline
\textit{Our} & 0.4827  & 0.6286 & 0.4160 & \textbf{0.5090} \\
\hline
\end{tabular}
\label{tab:urmp}
\end{table}

\section{Experiment}
\subsection{Baselines}
We compare our method with two baseline models, which share similar objectives with our work to generate conditioned body movement based on the given audio data.

\textbf{Audio to body dynamics}
Both our work and \cite{shlizerman2018audio} aim to generate body movement in music performance.
\cite{shlizerman2018audio} predicts reasonable playing movement based on piano and violin audio. Their model consists of 1-layer LSTM layer with time delay and 1-layer fully connected layer with dropout. We follow their setup and use MFCC feature as the input. It should be noted that while  
PCA is applied in \cite{shlizerman2018audio} to reduce the dimension in lower hand joints, PCA is not applicable to our task, since our task is to generate the full body motion, instead of only generating the hand motion. 
Their work takes the estimated 2-D arm and hand joint positions from video as the psudo-labels, whereas we extracts 15 body joints in 3-D space.

\textbf{Speech to gesture}
Another work in \cite{ginosar2019learning} aims 
to predict the speaker's gesture based on the input speech audio signal. Compared to our task, their predicted gesture motion are short segments ranging from 4-12 seconds, while our music pieces generally range from one to ten minutes. Convolutional U-net architecture is applied to their work, and a motion discriminator is introduced to eliminate single motion output. Although applying discriminator may increase the distance between the generated motion and ground truth(i.e, $L_1$ loss), the model is capable of producing more realistic motion. In this paper, we only take their model without discriminator as the baseline for comparison. 

\subsection{Evaluation metrics}
There is no standard way to measure the performance of a body movement generation system so far. To provide a comprehensive comparison among different methods, we propose a rich set of quantitative metrics to measure the overall distance between the skeletons and also the accuracy of bowing attack inference.

$\boldsymbol{L_1}$ \textbf{and PCK}
While $L_1$ distance is the objective function in the training stage, we also used it to evaluate the difference between generated motion and ground truth. Note that we report the results by averaging over 45-D joints and across all frames. Considering that the motion in right-hand wrist is much larger compared to other body joints, we calculate another $L_1$ hand loss for the 3-D wrist joint alone. 
The Percentage of Correct Keypoints (\textbf{PCK}) was applied to evaluate the generated gesture in speech in a prior work \cite{ginosar2019learning}, and we adapt PCK to 3-D coordinate in this paper. In the computation of PCK, a predicted keypoint is defined as correct if it falls within $\alpha\times \max(h,w,d)$ pixels of the ground truth keypoint, where $h$, $w$ and $d$ are the height, width and depth of the person bounding box, and we average the results using $\alpha=0.1$ and $\alpha=0.2$ as the final PCK score.

\textbf{Bowing attack accuracy}
The bowing attack indicates 
the time slot when the direction of the hand is changing. We first take both right-hand wrist joint sequences having length $L$ as $\hat{y}^{(rw)}$ and $y^{(rw)}$. 
Note that $\hat{y}^{(rw)}$ here only represents one coordinate of right-hand wrist joint. For each coordinate, We then compute the direction $D(i)$ for both sequences as:

\begin{equation}\label{eq:9}
D(i) =
  \begin{cases}
    1  & \quad \text{if } y^{(rw)}(i+1) - y^{(rw)}(i) > 0,\\
    0  & \quad \text{if } y^{(rw)}(i+1) - y^{(rw)}(i) \leq 0.
  \end{cases}
\end{equation}

Accordingly, we 
get the 
right-hand wrist joint direction for generated results $\hat{D}(i)$ and ground truth $D(i)$ respectively. Derived from the bowing direction $D(i)$, the bowing attack $A(i)$ at time $i$ would be set as 1 if the direction $D(i)$ is different from $D(i-1)$:
\begin{equation}
\label{eq:10}
A(i) =
  \begin{cases}
    1  & \quad \text{if } D(i) - D(i-1) \neq 0,\\
    0  & \quad \text{otherwise }.
  \end{cases}
\end{equation}

Finally, we compare the predicted bowing attacks $\hat{A}(i)$ and the ground truth ones $A(i)$. Additionally, we take a tolerance $\delta$, and set the ground truth as 1 in the range $[i-\delta, i+\delta]$ for a  
bowing attack located at time $i$, which suggests that the predicted bowing attack $\hat{A}(i)$ is a true positive (i.e. correct) prediction, if real bowing attack is located on the range $[i-\delta, i+\delta]$. Otherwise, it would be a false prediction. 
Notice that all the ground truth bowing attacks are only calculated once, which means that if all real bowing attacks near $\hat{A}(i)$ have been calculated before, then $\hat{A}(i)$ is a false prediction. Another previous work~\cite{liu2020body} also introduced bowing attack accuracy as an evaluation metric, and set the tolerance value as
$\delta=10$ (i.e, 0.333s). 
We consider that the tolerance should be more strict and set $\delta=3$ (i.e, 0.1s) in this paper. The F1-scores for bowing attack labels on axes x, y, z (width, height and depth) are calculated, and 
represented as Bowx, Bowy and Bowz, respectively, whereas the average of three bowing attack accuracy is shown as bow avg.

\textbf{Cosine similarity}
In this paper, our goal is not to generate identical 
playing movement as the ground truth, and the  cosine similarity is therefore a suitable measurement to evaluate the general trend of bowing movement. We compute cosine similarity for 3-D right-hand wrist joint between generated results and the ground truth, and then take the average over three coordinates across all frames.

It should be noted that the above evaluation metrics cannot measure the \emph{absolute} performance of body movement generation, since there is no standard and unique ground truth. Instead, the above evaluation metrics are to measure the \emph{consistency} between the generated results and a version of human performance.

\subsection{Quantitative results}

We 
compare our proposed method with two baselines~\cite{shlizerman2018audio}~\cite{ginosar2019learning} for the average performance over 14-fold test (as shown in Table~\ref{tab:quantitative results}). Three variants of the proposed methods are presented: First, {\it Unet1} represents the model with one single 
block (i.e. $M=1$, see Figure \ref{overview}) composed by U-net with self-attention. Second, {\it Unet1 + FFN + Refine Network} 
adds a 
position-wise feed forward network and a refine network after {\it Unet1}. The last one, {\it Unet2 + FFN + Refine Network}, adopts 
two U-net blocks ($M=2$) instead. 
The values reported in Table~\ref{tab:quantitative results} can be understood based on a reference measurement: the mean of right arm length in our dataset is 0.13. For example, an $L_1$ average values at 0.0391 mean that the average $L_1$ distance between ground truth and prediction is around 0.0391/0.13$\approx$30\% of the length of right arm. In addition, it should be noted that $L_1$ avg are generally 
smaller than $L_1$ hand avg, which is owing to the fact that 
joints on trunk mostly move with small quantity, 
whereas right-hand wrist exhibit obvious bowing motion covering wide moving range. 



It can be observed from the table that our 
model outperforms {\it A2B} both in $L_1$ and PCK, which indicates that our model applying 
U-net based network with self-attention mechanism 
can improve the performance for learning ground truth movement. Although {\it S2G} has competent performance for $L_1$ avg and PCK, our model boosts bowing attack accuracy more than 4\% compared to {\it S2G}. Also, after adding the position-wise feed forward network and the refined 
network, we get better performance in $L_1$ hand, bowing attack x, y, z and cosine similarity. This proves that the two components play a critical role for learning hand movement. Further, 
stacking two U-net blocks can increase bowing attck accuracy about 1\%. Overall, stacking two blocks of U-net and adding two components can achieve the 
best results in most metrics. 
This best model outperforms the baseline {\it A2B} model significantly in a two-tailed t-test ($p=8.21\times 10^{-8}$, d. f. $=250$). 

\subsection{Cross-dataset evaluation}
To explore if the methodology and 
designed process can adapt to other scenarios (e.g., different numbers of recorded joints, different positions such as standing or sitting, etc.), 
a cross-dataset evaluation is performed on 
the URMP dataset. The same process mentioned in Section 3 is applied to extract audio features and to estimate violinists' skeleton motion. 
However, the URMP dataset only contains 13 body joints, whereas 15 joints are 
extracted from our recorded videos. 
Considering the different skeleton layouts between two dataset, only the averaged bowing attack accuracy, and the accuracy on three directions are 
computed as illustrated in Table~\ref{tab:urmp}. 
Our method (i.e. \emph{Unet2 + FFN + Refine Network}) in Table~\ref{tab:urmp} 
represents the best model demonstrated in the quantitative results. It can be observed that our proposed method outperforms two baselines for bowing attack accuracy, and it is demonstrated that our model is well-adapted to different scenarios and datasets.

\begin{table}[t]
\caption{Comparison for baselines and the proposed model evaluated on audio input with varying speeds. `1x' means the original speed, `2x' means double speed, and so on.}
\begin{tabular}{|l|ccccc|}
\hline
             & 0.5x    & 0.75x   & 1x      & 1.5x    & 2x      \\ \hline\hline
\textit{A2B}~\cite{shlizerman2018audio} & 0.4024   & 0.4217 & 0.4357 & 0.4807 & 0.4971 \\
\textit{S2G}~\cite{ginosar2019learning} & 0.3591 & 0.3744 & 0.3921 & 0.4007 & 0.4111 \\ \hline
\textit{Our} & \textbf{0.4400}   & \textbf{0.4367} & \textbf{0.4656} & \textbf{0.4896} & \textbf{0.5182}\\
\hline
\end{tabular}
\label{tab:varied speed}
\end{table}

\subsection{Robustness test}
To test the robustness of our 
model to tempo variation, we compare the average bowing attack F1-scores on the same music pieces with different tempi. 
It is expected that 
the performance of a robust model should be invariant with diverse tempi. It should be noted that only the longest piece in the dataset is tested in this experiment, and all results shown in Table~\ref{tab:varied speed} are the Bow avg values only. 
As shown in Table~\ref{tab:varied speed}, our proposed model achieves better results compared to two baselines in five settings of tempo, which verifies the robustness of the proposed method with different performance tempi. 
The bowing attack accuracyis more likely to improve with faster tempo, since the prediction has a better chance to fall between 
the range $[i-\delta, i+\delta]$. 

\subsection{Subjective evaluation}
\begin{figure}[t]
\centering\includegraphics[width=3.5in]{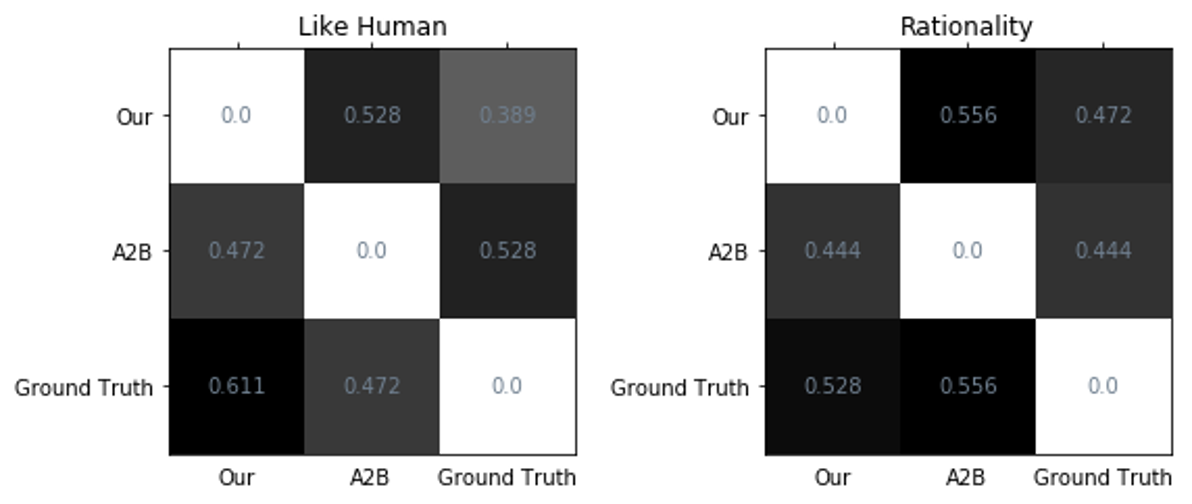}
\caption{The subjective evaluation for all participants. Left: The extent of playing movement being like human. Right: The rationality of playing movement.}\label{subjective_evaluation_all}
\end{figure}

\begin{figure}[]
\centering\includegraphics[width=3.5in]{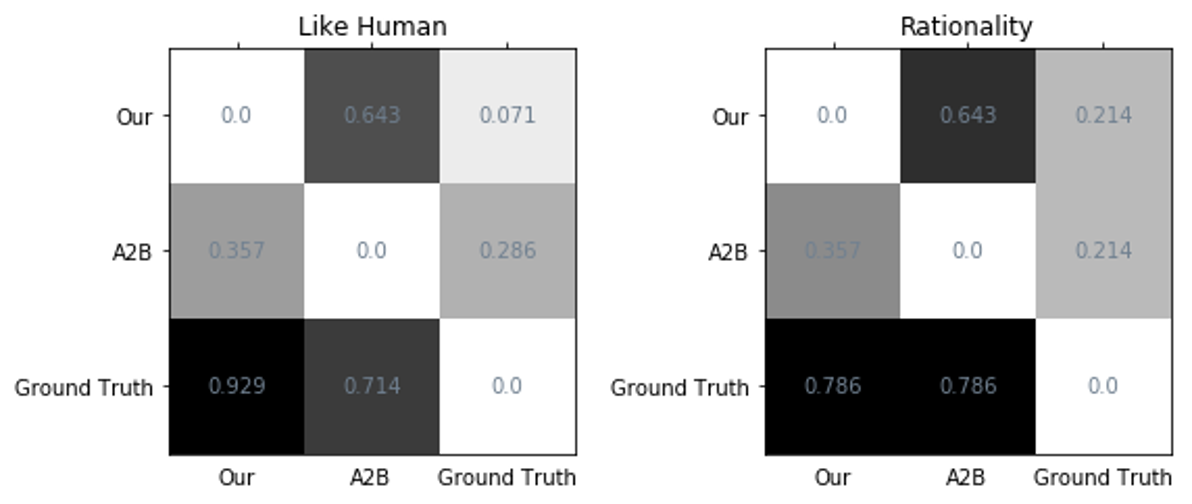}
\caption{The subjective evaluation only for the participants who have played violin. Left: The extent of playing movement being like human. Right: The rationality of playing movement.}\label{subjective_evaluation_vio}
\end{figure}

Since {\it A2B} shows better performance than {\it S2G}, we take only the ground truth, {\it A2B}, and our model for comparison in the subjective evaluation. The material for evaluation is 14 performances 
played by one randomly selected violinist (length of each performance = 94 seconds). 
The ground truth, the generated movements by {\it A2B} and our model are presented in a random order for per music piece.
And the participants are asked to rank 'the similarity level compared to human being's playing' and the 'rationality of the movement arrangement' among the three versions.
Within 36 participants in this evaluation, 38.9\% of participants have played violin, and 41.7\% have gotten music education or worked on music-related job. 
The results for all participants are shown in Figure~\ref{subjective_evaluation_all}, whereas the results specifically for the participants who have played violin are shown in Figure~\ref{subjective_evaluation_vio}. 

For the rationality, our results and the ground truth are much more reasonable than $A2B$, and the difference is more evident in Figure~\ref{subjective_evaluation_vio} compared to Figure~\ref{subjective_evaluation_all}. For the extent of being like human, the results are quite similar to the results of rationality in Figure~\ref{subjective_evaluation_vio}, whereas no obvious trend is observed in Figure~\ref{subjective_evaluation_all}. 
This result may be owing to the limitation that only the violinist's skeleton is included in the evaluation.
In the future work, we consider to incorporate the violin bow as a part of our architecture to generate more vivid animations. 

\subsection{Qualitative results}

\begin{figure}[t]
\centering\includegraphics[width=0.5\textwidth]{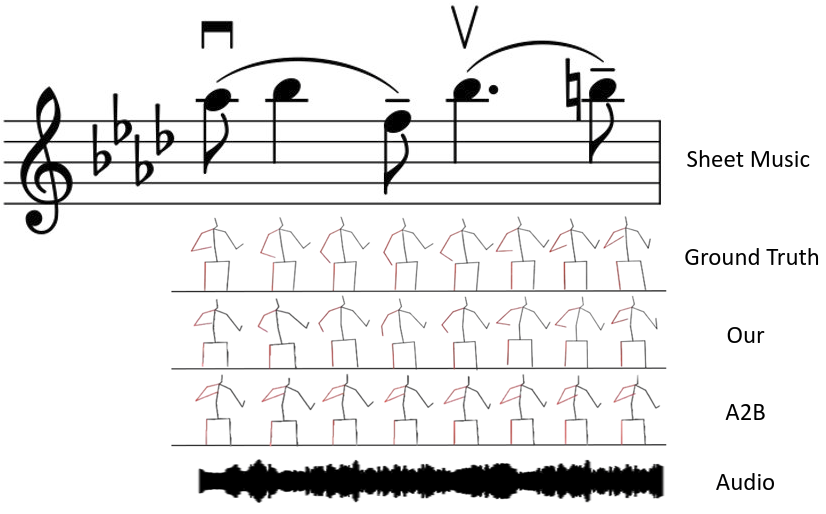}
\caption{Illustration of generated playing movement and ground truth with corresponding sheet music. $\sqcap$ and $\vee$ indicate down bow and up bow separately. Example selected from the 20th bar of a folk song \emph{Craving for the Spring Wind} composed by Teng Yu-Shian.}\label{qualitative evaluation2}
\end{figure}

For a more comprehensive demonstration of our result, we illustrate one example of the generated skeletons of the proposed method, the baseline method, and ground truth, as shown in Figure~\ref{qualitative evaluation2}. In this example, we choose one bar from one of the music pieces in the testing data and show the corresponding skeletons. Figure~\ref{qualitative evaluation2} clearly shows that the movements of ground truth skeletons are consistent to the down-bow and up-bow marks in the score.
It can be observed that the skeletons generated by the proposed model also exhibit consistent bowing direction in the right hand, while 
the skeletons generated by A2B do not show any changes of the bowing direction within this music segment.

\section{Conclusion}
\label{sec:conclusion}
In this paper, we have demonstrated a novel method for music-to-body movement generation in 3-D space. Different from previous studies, which merely apply conventional recurrent neural networks on this task, our model incorporates the U-net with self-attention mechanism to enrich the expressivity of skeleton motion. Also, we design a refinement network specifically for the right wrist to generate more reasonable bowing movements. Overall, our proposed model achieves promising results compared to baselines in quantitative evaluation, and the generated body movement sequences are perceived as reasonable arrangement in subjective evaluation, especially for participants with music expertise. Codes, data, and related materials are available at the project  link.\footnote{https://github.com/hsuankai/Temporally-Guided-Music-to-Body-Movement-Generation}

\begin{acks}
This work is supported by the Automatic Music Concert Animation (AMCA) project funded by the Institute of Information Science, Academia Sinica, Taiwan. The authors would also like to thank Yu-Fen Huang for editing this paper and discussions.
\end{acks}

\bibliographystyle{ACM-Reference-Format}
\bibliography{reference}

\end{document}